\documentstyle[aps,prl,twocolumn]{revtex}

\begin{document}
\author{V. V. Dobrovitski, M. I. Katsnelson, and B. N. Harmon}
\address{Ames Laboratory, Iowa State University, Ames, Iowa, 50011}
\title{Statistical coarse-graining as an approach to
multiscale problems in magnetism}

\draft

\maketitle

\begin{abstract}
Multiscale phenomena which include several processes 
occuring simultaneously at different length scales and exchanging
energy with each other, are widespread in 
magnetism. These phenomena often govern the magnetization
reversal dynamics, and their correct modeling
is important. In the present paper, we propose an 
approach to multiscale modeling of magnets, applying 
the ideas of coarse graining. 
We have analyzed the choice of the 
weighting function used in coarse graining, and propose
an optimal form for this function.
Simple tests provide evidence that this approach 
may be useful for modeling of realistic magnetic
systems.
\end{abstract}


A large number of phenomena taking place in magnets include
processes occuring simultaneously at different length scales. A
good example is the magnetization reversal in a macroscopic piece
of magnetic material possessing different kinds of defects,
voids, surfaces etc. The reversal starts by a nucleation of a
domain with the magnetization opposite to the initial direction.
As a rule, the nucleation happens near defects, where spins can
be frustrated. Here, the different length scales involved can be
clearly identified. First, there is the microscopic scale with a
characteristic length of the order of several interatomic
distances (several tens of angstroms), which corresponds to the
region of spin frustration and contains the microstructure in
the vicinity of the defect. Next, there is a ``micromagnetic''
length scale (of order of the domain wall width, several
thousands of angstroms) where the formation of the general
structure of the nucleus takes place. And, finally, the truly
macroscopic length scale (of order of several microns or even
millimeters), where the magnons created in the course of the
reversal propagate. These magnons play an important role in
energy transfer \cite{safonov} and sometimes can initiate the
magnetization reversal in other areas of the sample
\cite{avalanche}. A similar picture of several interacting length
scales appears in many situations, such as the breakthrough of a
domain wall pinned by a defect \cite{torredw}, influence of the
surface on the magnetic structure of the core of a magnetic
particle \cite{kodama} etc.

Processes of this type, called multiscale processes, are
receiving considerable attention nowadays. Along with very
interesting and rich physics, these are the very processes which
govern the switching behavior of magnetic systems (coercive
field, switching time, etc.), so that an adequate understanding
of multiscale phenomena is of paramount importance for
development and creation of new magnetic storage media.
Micromagnetic simulations can provide a realistic description of
the processes taking place at both micromagnetic and macroscopic
length scales. On the other hand, microscopic inhomogeneities 
require atomistic simulations (e.g., spin dynamics
modeling \cite{landau} is an adequate tool for materials where
spins are well localized at the sites of the crystalline lattice).
However, for the description of real systems, all the length
scales should be coupled, i.e. modeled simultaneously and
seamlessly, with the possibility of energy transfer between them.

A simple scheme, where the region of micromagnetic simulations is
just attached to the region of atomic spin simulations, does not
give a satisfactory description of the energy transfer between
the lengthscales. The problem is that micromagnetics does not
take into account atomistic degrees of freedom; even reduction of
the computational mesh in micromagnetic modeling down to atomic
scale does not describe the short-wavelength excitations
properly, contrary to spin dynamics simulations. The artificial
sharp boundary appearing between the micromagnetic and atomistic
regions leads to the unphysical scattering of magnons
transferring the energy from one region to the other. To couple
length scales correctly, some transition region between the
micromagnetic and atomic regions is necessary, which allows for
gradual exclusion of short-wavelength modes until they die out in
the regions far from the defect.

Similar difficulties arise in modeling dynamics of structural
defects in crystals (dislocation motion, crack propagation,
etc.). The coupling of lengthscales for this class of problems
has been extensively discussed in the literature 
\cite{kohlhoff,philips,rudd}. One of the most promising 
approaches to the
solution of this problem is coarse-grained molecular dynamics
(CGMD), developed in Ref.\ \cite{rudd}. In the present paper, we
apply the ideas of CGMD to magnetic materials modeling. We
propose a computational scheme which employs basic concepts of
nonequilibrium statistical mechanics to couple micromagnetism and
the dynamic modeling of classical spins.  We identify the key
problems arising in the course of implementation of this scheme
and their possible solutions.

To explain the idea of the method, 
let us consider a ferromagnet where some magnetic inhomogeneity
(defect, interface, etc.) is present. We describe it as a 
system of classical magnetic moments
${\bf M}_{\mu}$ of fixed length $M=|{\bf M}_{\mu}|$, located 
at the $\mu$-th site of the crystalline lattice (greek indices
enumerate the lattice sites).
We assume that the system is described by the Hamilton function
\begin{eqnarray}
\label{ham}
{\cal H} &=& {\cal H}^0 + {\cal V},\\ \nonumber
{\cal H}^0 &=& \sum_{\mu} {\cal H}_{\mu}^0 = \sum_{\mu,\nu} J_{\mu\nu} 
  {\bf M}_{\mu} {\bf M}_{\nu},
\end{eqnarray}
where ${\cal H}^0$ describes the isotropic exchange interaction,
while ${\cal V} \ll {\cal H}^0$ represents
all the other, much weaker interactions present in the system, 
and $J_{\mu\nu}$ is the exchange integral between the sites $\mu$
and $\nu$. 
Let us focus on the region far from the defect, where 
the amplitudes of short-wavelength excitations are small
(the region where their amplitudes are not small should be treated
completely atomistically). Suppose, a computational grid
is defined in this region, so that the magnetic state is described
with required precision by defining the magnetization direction
at the grid nodes: ${\bf M}_j=(M^x_j, M^y_j, M^z_j)$ at 
the $j$-th node (below, latin indices enumerate the nodes of
the mesh). These data represent the necessary number
of large-scale degrees of freedom, and exact modeling
of all the atomic-scale modes would be too excessive
(and too expensive resoursewise). However, the atomic-scale modes
can not be excluded completely, since their cumulative
effect (dissipation, energy transfer, etc.) is not 
negligible, and should be taken into account.

This problem can be solved by invoking statistical physics, 
i.e. under the assumption of ergodicity the exact dynamics
of short-scale modes can be replaced by their 
statistical distribution. For many relevant situations it has been
shown \cite{bogoliubov} that the short-scale modes relax 
almost immediately to the state of local quasiequilibrium 
determined uniquely by large-scale modes,
so that knowledge of only the large-scale parameters 
(``gross variables'') determines
the dynamics of the system with necessary rigor.
This description of the system, where only large-scale modes
are essential (while short-scale modes are ``slaved''
due to the requirement of the local equilibrium),
is often referred to as the {\it coarse-grained\/} description.
The theory of local quasiequilibrium states has been
developed in the 1950s-1960s, and a number of approaches exist
\cite{zubarev,green,zwanzig,mori}. In the following, we will 
use the convenient modification of the nonequilibrium
statistical operator (NSO) method \cite{bz}. For simplicity, 
we do not consider the dissipation processes and quantum spin 
effects, but in principle they can also be included following
Ref.\ \cite{bz}.

It is convenient to introduce new dynamic variables $\alpha_{1,\mu}$ and
$\alpha_{2,\mu}$ for the magnetic moments ${\bf M}_{\mu}$
according to the following relations:
\begin{eqnarray}
M^x_{\mu} &=& \alpha_{1,\mu} \sqrt{2M - \alpha_{1,\mu}^2 - 
  \alpha_{2,\mu}^2},\\
 \nonumber
M^y_{\mu} &=& \alpha_{2,\mu} \sqrt{2M - \alpha_{1,\mu}^2 - 
  \alpha_{2,\mu}^2},\\
 \nonumber
M^z_{\mu} &=& M - \alpha_{1,\mu}^2 - \alpha_{2,\mu}^2.
\end{eqnarray}
These variables can be expressed via conventional polar $\theta_{\mu}$
and asimuthal $\phi_{\mu}$ angles as 
\begin{eqnarray}
\alpha_{1,\mu} &=& \sqrt{2M} \sin{(\theta_{\mu}/2)} \cos{\phi_{\mu}},\\
  \nonumber
\alpha_{2,\mu} &=& \sqrt{2M} \sin{(\theta_{\mu}/2)} \sin{\phi_{\mu}}.
\end{eqnarray}
It can be checked that the variables $\alpha_{1,\mu}$ and 
$\alpha_{2,\mu}$ are canonically 
conjugated, so that Hamiltonian formalism can
be used, and a computational scheme preserving the
simplectic structure can be employed. 

The conventional way to develop a coarse-grained description of the
system is to introduce the large-scale variables $\alpha_j$ as
averages of $\alpha_{\mu}$:
\begin{equation}
\alpha_{1,j} = \sum_{\mu} f_{\mu,j} \alpha_{1,\mu},\qquad 
\alpha_{2,j} = \sum_{\mu} f_{\mu,j} \alpha_{2,\mu},
\end{equation}
where $f_{\mu,j}$ is an appropriate weighting function,
satisfying the normalization condition $\sum_{\mu} f_{\mu, j} = 1$,
and localized near the node $j$; proper choice of the weighting 
function $f_{\mu,j}$ will be discussed below in more detail.

Averaging is a natural way to introduce gross variables
for a ferromagnet, where the local equilibrium
is governed by the Hamiltonian ${\cal H}^0$, so that
all the magnetic moments ${\bf M}_{\mu}$
near the node $j$ are almost parallel to ${\bf M}_j$.
Introduced in this way, $\alpha_{(1,2),j}$ constitute
collective variables of the system \cite{bohm} changing
slowly with respect to quickly relaxing short-scale modes,
so they can be considered as quasi-integrals of
motion (for more detailed discussion see Refs.\ 
\cite{bogoliubov,bz}). Following standard procedures
of statistical physics, the integrals of motion are
included into the distribution function via Legendre 
transformations, i.e. we introduce Legendre multipliers
$F_j$ and $G_j$ corresponding to the node variables 
$\alpha_{1,j}$ and $\alpha_{2,j}$, so that the distribution 
function can be written as
\begin{eqnarray}
\label{dist}
\rho &=& Q^{-1} \exp{\Bigl( -\sum_{\mu} \beta_{\mu} {\cal H}^0_{\mu} 
  - \sum_j F_j \sum_{\mu} \beta_{\mu} f_{\mu,j} \alpha_{1,\mu}}\\
 \nonumber
  && - \sum_j G_j \sum_{\mu} \beta_{\mu} f_{\mu,j} \alpha_{2,\mu}
  \Bigr),\\ \nonumber
Q &=& \int \exp{\Bigl( -\sum_{\mu} \beta_{\mu} {\cal H}^0_{\mu} - 
  \sum_j F_j \sum_{\mu} \beta_{\mu} f_{\mu,j} \alpha_{1,\mu}}\\
 \nonumber
  && - \sum_j G_j \sum_{\mu} \beta_{\mu} f_{\mu,j} \alpha_{2,\mu}
  \Bigr)\ \prod_{\mu} \frac{d\alpha_{1,\mu}\, d\alpha_{2,\mu}}{2\pi M},
\end{eqnarray}
where $\beta_{\mu}=1/(k_B T_{\mu})$, $k_B$ is Boltzmann's constant,
$T_{\mu}$ is the spin temperature at the $\mu$-th
site, and $Q$, as can be seen, is the normalization factor, analogous
to the statistical sum of a canonical ensemble in the 
equilibrium case. Note that smooth variations of the temperature
across the sample can be taken into account (temperature is 
a Legendre multiplier for the integral of energy), but to make the
consideration simpler, we neglect it, assuming constant temperature,
small enough to satisfy the condition $\beta J_{\mu\nu}\gg 1$.

The variables $F_j$ and $G_j$ can be considered as parameters
of a local fictitious field acting on the moments ${\bf M}_{\mu}$.
Values of the parameters $F_j$ and $G_j$ should be
chosen in such a way that the resulting equilibrium values 
$\alpha_{(1,2),\mu}$
averaged with the weighting function $f_{\mu,j}$ give the required
values of $\alpha_{(1,2),j}$. It is easy to see from Eq.\ 
\ref{dist} that the values $F^0_j$, $G^0_j$ producing the 
required values $\alpha_{(1,2),j}$ are determined from the equations
\begin{eqnarray}
\label{pars}
\left.\frac{\partial {\cal F}}{\partial F_j}\right|_{F^0_j,G^0_j}
  &=& \alpha_{1,j},\quad
  \left.\frac{\partial {\cal F}}{\partial G_j}\right|_{F^0_j,G^0_j}
  = \alpha_{2,j}, \\
 \nonumber
{\cal F}(F_j, G_j) &=& -\beta^{-1} \ln{Q}
\end{eqnarray}
where $\cal F$ is an analog of the Gibbs' free energy function.

Now, having all the information at hand, we can use averaging
to get the equations of motion for the coarse-grained variables
$\alpha_{(1,2),j}$. The underlying dynamics of the microscopic
variables $\alpha_{(1,2),\mu}$ is Hamiltonian, i.e.
\begin{equation}
\dot\alpha_{1,\mu}= -\frac{\partial {\cal H}}{\partial\alpha_{2,\mu}},
  \quad
  \dot\alpha_{2,\mu}= \frac{\partial {\cal H}}{\partial\alpha_{1,\mu}},
\end{equation}
where ${\cal H}$ is the Hamilton function (\ref{ham}).
Using the distribution function (\ref{dist})
with the values $F^0_j$ and $G^0_j$ determined from Eq.\
\ref{pars}, we obtain:
\begin{eqnarray}
\label{eqmot}
\dot\alpha_{1,j} &=& -\sum_{\mu} f_{\mu,j} \Big\langle
  \frac{\partial {\cal V}}{\partial\alpha_{2,\mu}} \Big\rangle
  -\sum_{\mu} f_{\mu,j} \sum_k f_{\mu,k} G^0_k, \\
 \nonumber
\dot\alpha_{2,j} &=& \sum_{\mu} f_{\mu,j} \Big\langle
  \frac{\partial {\cal V}}{\partial\alpha_{1,\mu}} \Big\rangle
  +\sum_{\mu} f_{\mu,j} \sum_k f_{\mu,k} F^0_k,
\end{eqnarray}
where $\langle\dots\rangle$ means averaging with the distribution
(\ref{dist}). Note that the equations of motion are nonlocal
over the node indices even if only local exchange interactions are
present in the system; this important feature 
is totally missing in the micromagnetic description.

Direct implementation of the scheme presented above can be
rather expensive computationally. To make the
problem easier, we can take into account that these
computations are to be performed only inside the relatively 
narrow ``belt'' between micromagnetic and atomic-scale regions
(see Fig.\ \ref{fig1}),
where departures of magnetization from equilibrium
are already small (otherwise, atomic-scale
simulations should be used). If the $z$-axis of 
the co-ordinate frame is aligned with the 
equilibrium direction of magnetization the values of 
$\alpha_{(1,2),\mu}$ are small,
and the Hamiltonian ${\cal H}^0$ can be expanded in
terms of $\alpha_{(1,2),\mu}$ up to second order:
\begin{equation}
\label{hamlin}
{\cal H}^0 = \frac 12\sum_{u,v=1,2} \sum_{\mu,\nu} D_{u\mu, v\nu}\,
  \alpha_{u,\mu}\, \alpha_{v,\nu},
\end{equation}
where the indices $u, v=1,2$ denote the variables $\alpha_1$
and $\alpha_2$. 
In this case, the distribution (\ref{dist}) is Gaussian,
so that the integral $Q$ and the function $\cal F$ in (\ref{pars})
can be calculated exactly \cite{schulman}. As a result,
Eq.\ \ref{pars} determining the values of generalized torques
$F^0_j$ and $G^0_j$ can be written as a set of linear equations:
\begin{equation}
\label{parslin}
\alpha_{u,j} = \sum_k \sum_{v=1,2} T^0_{v,k} \sum_{\mu,\nu} 
  f_{\mu,j} D^{-1}_{u\mu, v\nu} f_{\nu,k},
\end{equation}
where $D^{-1}_{u\mu, v\nu}$ is the inverse of the dynamic matrix
$D_{u\mu, v\nu}$ in the linearized exchange Hamiltonian (\ref{hamlin}),
and we used the vector $T^0_{v,k}$ ($v=1,2$) to denote both
$F^0_k$ and $G^0_k$ as follows: $T^0_{1,k}\equiv F^0_k$, and 
$T^0_{2,k}\equiv G^0_k$. 

Hence, the problem of dynamic coupling of length scales
is reduced to two linear problems, Eqs.\ \ref{parslin} and 
\ref{eqmot}. Note that the rotation of the co-ordinate
frame which brings the $z$-axis into coincidence with
the equilibrium direction of magnetization makes the
dynamic matrix independent of $\alpha_{(1,2),j}$, so that
its inverse can be calculated once and stored for further
references. 
However, there is a subtlety in inverting $D_{u\mu, v\nu}$:
this matrix, being determined only by the isotropic exchange 
interactions, has a zero eigenvalue corresponding to a
shift of all $\alpha_{(1,2),\mu}$ by the same value, or,
in other words, the dynamic matrix has an eigenvector
$d^0 = (1,1,\dots,1)$ corresponding to the zero eigenvalue. It 
reflects the fact that the exchange Hamiltonian (\ref{hamlin})
is invariant with respect to rotation of the system
as a whole. Thus, when inverting numerically the dynamic matrix, 
a component corresponding to the zero eigenvector $d^0$
should be excluded. 

In the present form, the essense of coarse-graining
becomes especially clear. Imagine that, applying some
fictitious torques $F_j$ and $G_j$ we brought the
system into such a state that the equilibrium values
of the large-scale variables are $\alpha_{(1,2),j}$.
Then, the atomic magnetic moments ${\bf M}_{\mu}$ (i.e.,
the microscopic variables $\alpha_{(1,2),\mu}$) move
in such a way that, after the stage of quick relaxation 
is finished, their positions minimize the total energy
of the system with respect to the constraints imposed
by the torques $F_j$, $G_j$.

The last but not least problem is an appropriate choice of
the weighting function $f_{\mu,j}$: it can be verified
that an arbitrary function does not automatically give 
meaningful results. To study this question, let us inspect 
closely the basic idea of the coarse-grained description. 
For a general system consisting of $N$ microscopic moments,
an exact description of the system's dynamics requires
knowledge of {\it all\/} $2N$ microscopic canonical variables
$\alpha_{(1,2),\mu}$. However, we expect that under
certain conditions (which are yet to be formulated), the system
can be described with reasonable precision using the much
smaller set of $2L$ gross variables $\alpha_{(1,2),j}$.
In other words, we expect that under certain approximations,
we can define such a set of variables $\alpha_{(1,2),j}$ 
that allows specification of {\it all\/} microscopic variables 
$\alpha_{(1,2),\mu}$ with sufficient precision. In particular,
it can be shown \cite{rudd} that, for a given weighting function 
$f_{\mu,j}$, an optimal (in the least-square sense)
restoration of microvariables is achieved via linear 
transformation 
\begin{equation}
\label{rest}
\alpha_{u,\mu} = \sum_j N_{j, \mu} \alpha_{u,j},
\end{equation}
where 
$N_{j, \mu} = \sum_k f_{\mu, k} (\sum_{\nu} f_{\nu,j} f_{\nu, k})^{-1}$.
Thus, the problem is to find such a function $f_{\mu, j}$ 
which would make the restoration (\ref{rest}) as accurate as
possible.

In the coarse-grained region, where the linearized 
Hamiltonian (\ref{hamlin}) can be used, the system's 
dynamics can be represented as an
independent motion of different normal modes (eigenvectors of
the Hamiltonian (\ref{hamlin})), and the
choice of the $2L$ gross variables becomes obvious: they should
be amplitudes of the normal modes corresponding to lowest 
$L$ eigenfrequencies. This choice gives an almost complete description
of the system provided that the amplitudes of all the other
modes, which are excluded from consideration,
are much smaller. The frequencies of the
excluded modes are much larger, so that their dynamics
is much faster, and they can relax to local equilibrium
quickly in comparison with the adiabatically slowly
varying gross variables. 

This choice of the gross variables is in correspondence with the
dynamical approach \cite{bogoliubov} to nonequilibrium
statistical mechanics. When considering motion of the system
subjected to some small rapidly varying perturbation (provided,
in our case, by the fast excluded modes), the 
well-known basic problem is to exclude from the solution
so-called secular terms, which appear due to entanglement of slow
and fast motions in the system. After slow and fast modes are
properly separated, the standard procedure of averaging can be
performed. The use of the lowest-frequency normal modes as gross
variables allows elimination of the secular terms in the 
coarse-grained equations of motion. An analogous procedure can be
identifyed also in the approaches of Zwanzig \cite{zwanzig} and 
Mori \cite{mori}.

Thus, one possible recipe is to use the weighting function
$f_{\mu, j} = d^j_{\mu}$,
where $d^j$ is one of the lowest-frequency eigenvectors of
the dynamic matrix. 
In real calculations it could be inconvenient to work
with delocalized collective modes. It can be shown,
that an equally accurate coarse-graining
can be achieved with any orthonormal combination of the
eigenvectors $d^j$, i.e. $f_{\mu, j} = \sum_k C^j_k d^k_{\mu}$
is an equally good weighting function if
\begin{equation}
\label{cs}
\sum_k C^j_k C^{j'}_k = \delta_{j,j'}
\quad
\sum_j C^j_k C^j_{k'} = \delta_{k,k'}.
\end{equation}
The function $f_{\mu, j}$ can be made well-localized
using an appropriate set of $C^j_k$. 

As a specific example, we performed coarse-grained modeling
of magnons in a 1-D ferromagnetic chain consisting of $N$ 
classical magnetic moments with
nearest-neighbor and next-nearest-neighbor exchange
interactions. The corresponding Hamilton function is
\begin{eqnarray}
{\cal H} &=& \sum_{\mu} J^0_{\mu} {\bf M}_{\mu} ({\bf M}_{\mu+1}
  + {\bf M}_{\mu-1}) \\ \nonumber
  &&+ \gamma J^0_{\mu} {\bf M}_{\mu} ({\bf M}_{\mu+2}
  + {\bf M}_{\mu-2}),
\end{eqnarray}
and periodic boundary conditions are used. A computational 
mesh is imposed, consisting of
$L$ nodes, and two gross variables $\alpha_{1,j}$ and
$\alpha_{2,j}$ are attributed to each of $L$ nodes.
These two gross variables provide a coarse-grained description
of a block containing $n=N/L$ individual moments.

Several weighting functions have been used. One possible
choice, giving an {\it exact} magnon spectrum, is
\begin{equation}
f^{(0)}_{\mu, j} = \frac 1N \, \frac{\sin{\pi(\mu-j n)/n}}
  {\tan{\pi(\mu-j n)/N}},
\end{equation}
which corresponds to an orthonormal combination of exact
normal modes with coefficients 
$C^j_k = (1/\sqrt{N}) \exp{(-{\rm i}jkn)}$ where
${\rm i}=\sqrt{-1}$ (so it is 
a discrete analog of the Nyquist-Shannon uniform sampling).
However, this function, due to its long tails, is not
convenient for computations, and we tested its rescaled
cutoff modification:
\begin{eqnarray}
f^{(1)}_{\mu,j} =& A f^0_{\mu,j},&\quad  \mu -jn\le n,\\ \nonumber
f^{(1)}_{\mu,j} =& 0,&\quad \mu -jn > n,
\end{eqnarray}
where $A$ is the normalization factor necessary to satisfy the
condition $\sum_{\mu} f_{\mu, j}=1$.
A second form of the weighting function
\begin{eqnarray}
f^{(2)}_{\mu,j} =& 1/n, &\quad \mu -jn\le n,\\ \nonumber
f^{(2)}_{\mu,j} =& 0,&\quad \mu -jn > n,
\end{eqnarray}
although very far from optimal, can be used for
crude semi-qualitative computations, so we also
tested its performance.

The results of our tests, the dispersion curves
$\omega (k)$ for magnons with different wave vectors
$k$, and their group velocities $c(k)=d\omega / d k$, are
calculated with different weighting functions, as
shown in Fig.\ \ref{fig2}. Dispersion curves $\omega (k)$
describe the magnon dynamics in the chain, while
the group velocity curve $c(k)$ characterizes the propagation
of magnons (the latter should be tested separately since
a good approximation for $\omega (k)$ does not necessarily
imply a good approximation of its derivative).
For comparison, exact curves
are presented, along with the results of micromagnetic
calculations. The data points for $k=0$ and $k=2\pi / n$ are
not shown because the dynamic matrix formally has a singularity
at these values of wave vector. For very long-wavelength
magnons all types of modeling work rather well, but
for shorter wave lengths the differences are large.
The coarse-grained description
is better than the micromagnetic even for the worst
function $f^{(2)}_{\mu, j}$. For the appropriately
chosen weighting function $f^{(1)}_{\mu, j}$, in spite
of the cutoff, the difference with the exact solution
is very small, even at maximal allowed wave vector values.

Summarizing, we propose an approach to modeling of multiscale
processes in magnets, applying the ideas of coarse grained
molecular dynamics \cite{rudd} to magnetic modeling.
The scheme proposed employs basic concepts of nonequilibrium 
statistical mechanics to couple lengthscales. We discussed
possible implementation of this approach, paying
particular attention to the problem of the correct choice 
of the weighting function used in coarse graining.
Simple tests verify our conclusions and evidence that
this approach can be applicable to larger and more complicated
systems.

Authors would like to thank R.\ Rudd, J.\ Morris, O.\ N.\ Mryasov, 
R.\ Sabiryanov, and T.\ Schulthess for helpful discussions.
This work was partially carried out at the Ames Laboratory, which 
is operated for the U.\ S.\ Department of Energy by Iowa State 
University under Contract No.\ W-7405-82 and was supported by 
the Director of the Office of Science, Office of Basic Energy Research
of the U.\ S.\ Department of Energy. This work was partially supported 
by Russian Foundation for Basic Research, grant 98-02-16219.

\begin{figure}
\caption{In the vicinity of the defect (the defect is depicted 
as a dashed circle in the center), individual magnetic moments 
can depart significantly from ideal ferromagnetic order, 
because of lack of co-ordination, and due to changes in 
exchange and anisotropy constants. This region can be
described using atomistic modeling of individual moments.
In the region far from the defect, spins are almost parallel
to each other, and a micromagnetic description is 
appropriate. To couple these regions correctly, a transition
region is included, where the coarse grained description is
used.}
\label{fig1}
\end{figure}

\begin{figure}
\caption{Calculated dispersion curves $\omega (k)$ (above) for 
magnons with different wave vectors $k$, and their group velocities
$c(k)$ (below). Wave vector values $k$ are shown in units of $k_0=2\pi/n$.
Calculations have been done for the chain consisting of
$N=1024$ individual moments with the mesh containing
$L=32$ nodes. The value of the next-nearest exchange integral
is $J=1$ and the ratio $\gamma=0.2$ of 
nearest-neighbor to next-nearest-neighbor exchanges are used.
The results for micromagnetic modeling (MM), and for two versions
of coarse graining with different weighting functions (CG~1 for 
$f^{(1)}_{\mu,j}$ and CG~2 for $f^{(2)}_{\mu,j}$) are presented. 
Exact curves are shown for comparison.}
\label{fig2}
\end{figure}

\end{document}